# Bimodal fission in the Skyrme-Hartree-Fock approach*

A. STASZCZAK$^a$, J. DOBACZEWSKI$^{b,c,d}$, W. NAZAREWICZ$^{b,c,d}$

$^a$Institute of Physics, Maria Curie-Skłodowska University,
pl. M. Curie-Skłodowskiej 1, 20-031 Lublin, Poland
$^b$Department of Physics, University of Tennessee,
Knoxville, TN 37996, Knoxville, USA
$^c$Physics Division, Oak Ridge National Laboratory, P.O. Box 2008,
Oak Ridge, TN 37831, USA
$^d$Institute of Theoretical Physics, Warsaw University,
ul. Hoża 69, Warsaw, Poland

Spontaneous-fission properties of $^{256}$Fm, $^{258}$Fm, and $^{260}$Fm isotopes are studied within the Skyrme-Hartree-Fock+BCS framework. In the particle-hole channel we take the Skyrme SkM* effective force, while in the particle-particle channel we employ the seniority pairing interaction. Three static fission paths for all investigated heavy fermium isotopes are found. The analysis of these fission modes allows to describe observed asymmetric fission of $^{256}$Fm, as well as bimodal fission of $^{258}$Fm and symmetric fission in $^{260}$Fm.

## 1. Introduction

A phenomenon known as bimodal fission is related to the remarkable properties of spontaneous fission observed in several fermium and transfermium nuclei, e.g., $^{258,259}$Fm, $^{259,260}$Md, and $^{258,262}$No [1, 2, 3, 4]. In these systems, a sharp transition takes place from an asymmetric mass division in, e.g., $^{256}$Fm and $^{256}$No to a symmetric split in, e.g., $^{258}$Fm and $^{258}$No. Furthermore, the total kinetic energy (TKE) distribution of the fission fragments appears to be composed of two Gaussians with the maxima near 200 and 233 MeV. It was postulated (Refs. [3, 5, 6, 7]) that the higher-energy fission mode corresponds to a scission configuration associated with two touching nearly spherical fragments, with the maximum of Coulomb repulsion, whereas the lower-energy mode can be associated with more elongated fragments. Moreover, the higher-energy mode consistently produces narrow and symmetric mass distributions, while the mass distributions of fragments with lower TKEs are much broader and sometimes asymmetric [3].

---

* Presented at the Zakopane Conference on Nuclear Physics, September 4-10, 2006, Zakopane, Poland.

In this work we discuss total binding energies and mass hexadecapole moments calculated along the static fission paths of $^{256}$Fm, $^{258}$Fm, and $^{260}$Fm. In Ref. [8] we studied the associated collective inertia. Here, the main focus is on differences in spontaneous fission properties found in this region of heavy fermium isotopes.

## 2. The model

The calculations are carried out within the self-consistent constrained Skyrme-Hartree-Fock+BCS (SHF+BCS) framework. The effective Skyrme force SkM* [9] is used in the particle-hole channel, whereas a seniority pairing force is taken in the particle-particle channel to describe nuclear superfluidity. The seniority pairing force is treated within the BCS procedure, with the strength parameters defined as in Ref. [10]:

$$\begin{aligned} G^n &= [19.3 - 0.084\,(N-Z)]\,/A\,, \\ G^p &= [13.3 + 0.217\,(N-Z)]\,/A\,, \end{aligned} \quad (1)$$

and additionally scaled by

$$\tilde{G}^{n/p} = f_{n/p} G^{n/p}. \quad (2)$$

The scaling factors of Eq. (2), $f_n = 1.28$ and $f_p = 1.11$, were adjusted to reproduce the experimental [11] neutron ($\Delta_n = 0.696$ MeV) and proton ($\Delta_p = 0.803$ MeV) pairing gaps in $^{252}$Fm. The pairing-active space consisted of the lowest $Z$ ($N$) proton (neutron) single-particle states.

The self-consistent HF+BCS equations were solved using the code HFODD (v.2.25b) [12, 13, 14] that employs the Cartesian 3D deformed harmonic-oscillator finite basis. In the calculations, we took the lowest 1140 single-particle states for the basis. This corresponds to 17 spherical oscillator shells.

## 3. Fission barriers of $^{256}$Fm, $^{258}$Fm, and $^{260}$Fm isotopes

The results of our calculations are presented in Figs. 1, 2, and 3, which display the total binding energy ($E^{\text{tot}}$) and mass hexadecapole moment ($Q_{40}$) calculated along the static fission paths of $^{256}$Fm, $^{258}$Fm, and $^{260}$Fm, respectively. The fission paths were computed with a mass quadrupole moment ($Q_{20}$) used as a driving coordinate (constraint). Our study covers prolate shapes with $Q_{20} = 0 \div 550$ b.

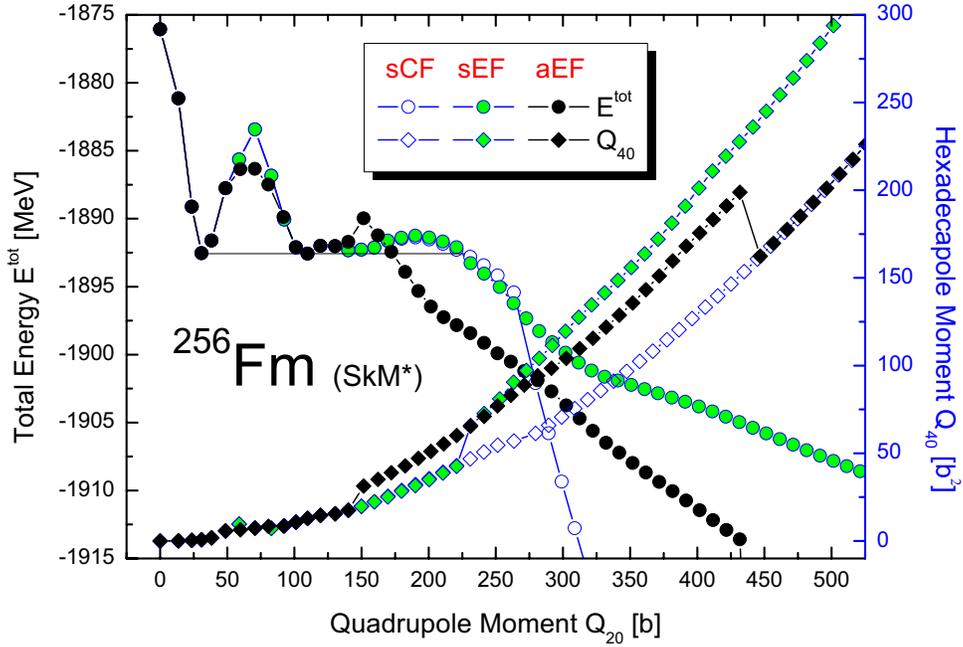

Fig. 1. The total binding energy $E^{\text{tot}}$ (left axis) and mass hexadecapole moment $Q_{40}$ (right axis) along the calculated symmetric compact (sCF), symmetric elongated (sEF), and asymmetric elongated (aEF) fission paths in $^{256}$Fm. Differences in $E^{\text{tot}}$ shown in the vicinity of the first (inner) barrier illustrate the effect of triaxiality.

A similar pattern of static fission trajectories was found for all investigated fermium isotopes. Beyond the region of the first fission barrier, at $Q_{20} \approx 150$ b, a reflection-asymmetric path corresponding to elongated fragments (aEF) branches away from the symmetric valley. At $Q_{20} \approx 225$ b, a reflection-symmetric path splits into two branches: one corresponding to a division into nearly spherical (compact) fragments (sCF) and the second branch corresponding to elongated fragments (sEF). It is worth mentioning that the identical pattern of static fission paths in $^{258}$Fm has been found recently in Ref. [15] within the same SHF+BCS framework, except that the axial-symmetry was enforced therein.

Furthermore, it appears that at $Q_{20} \approx 430$ b for $^{256}$Fm and at $Q_{20} \approx 470$ b for $^{258,260}$Fm, the calculated asymmetric aEF solution becomes unsta-ble and it *falls back* into the symmetric sCF one. For $^{256}$Fm, at this transition point the energy of the sCF configuration is equal to about $-1950$ MeV, and is not shown in Fig. 1 in order not to extend the scale of the figure too much (see, however, Fig. 5 of Ref. [8]), and similarly holds for $^{258,260}$Fm. All these calculated points of instability correspond to scission configurations of fragments in the asymmetric channel. The analysis of nuclear shapes at these points [16] shows that one of the fragments is elongated while the other one is close to a sphere. This observation is in agreement with results obtained in Ref. [17] within the Hartree-Fock-Bogoliubov approach with the D1S Gogny interaction.

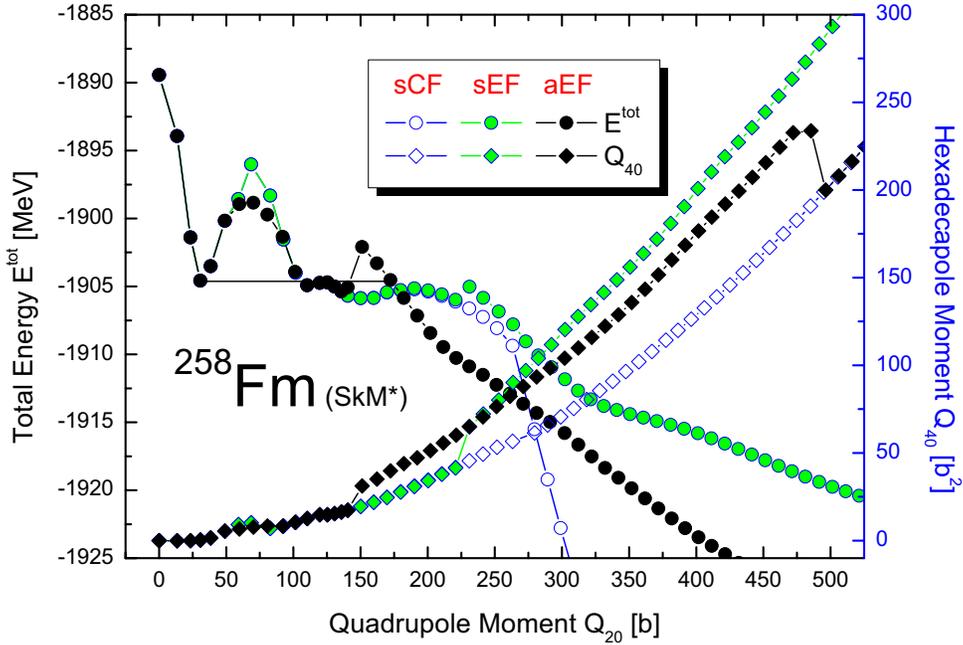

Fig. 2. Similar to Fig. 1 except for $^{258}$Fm.

In the symmetric channels, the scission configuration for the sCF path (with two nearly spherical touching fragments) appears at $Q_{20} \approx 260$ b (in all the investigated fermium isotopes), whereas for the sEF path, the scission configuration (with very elongated fragments) was reached at $Q_{20} \approx 810$ b only in $^{258}$Fm.

Although the general pattern of static fission paths is fairly similar for the three fermium isotopes, the analysis of the binding energy along these paths reveals significant differences. In $^{256}$Fm, for instance, the outer barrier at $Q_{20} \approx 150$ b is very narrow along the asymmetric path aEF, whereas it is predicted to be broad for the two symmetric paths (sCF and sEF), and the saddle point is moved to larger deformations, $Q_{20} \approx 200$ b. While the quantitative analysis of the competition between different fission valleys obviously requires the proper treatment of fission dynamics, the calculated pattern for $^{256}$Fm shown in Fig. 1 is strongly indicative of the favored character of the asymmetric path aEF for this isotope.

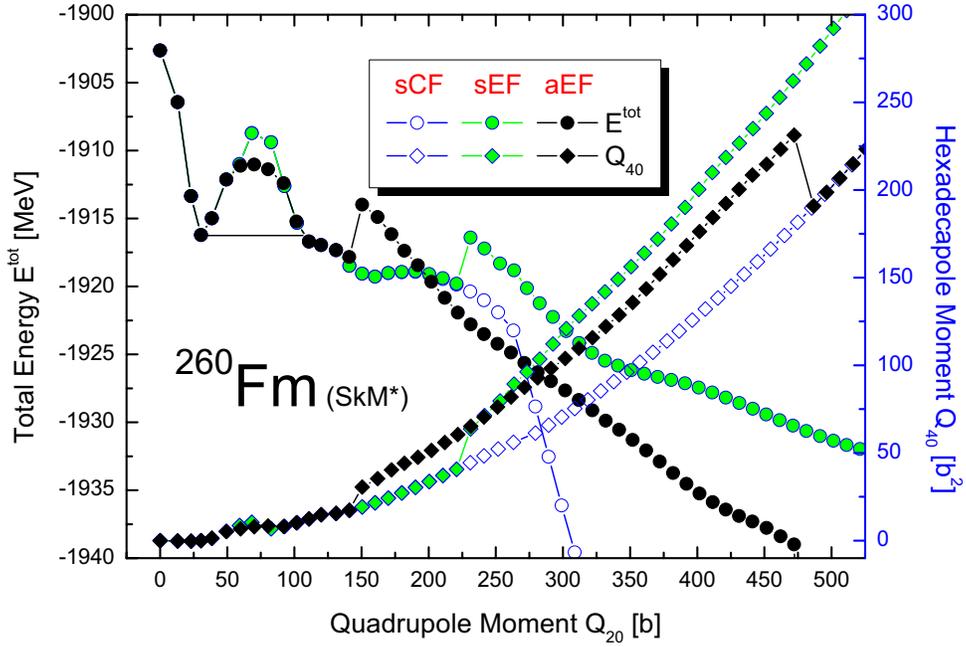

Fig. 3. Similar to Fig. 1 except for $^{260}$Fm.

In stark contrast to the situation in $^{256}$Fm, both symmetric paths are open for $^{258}$Fm, due to the disappearance of the outer fission barrier in sCF and sEF (Fig. 2). The sCF and sEF paths can be associated with the higher- and lower-TKE modes of the bimodal fission, respectively. Moreover, the less favorable aEF path may yield a small asymmetric contribution to the mass distribution of events with lower TKEs, as was postulated in Ref. [3].

In the case of $^{260}$Fm (Fig. 3), we find that there is no outer potential barrier along the sCF trajectory, and the sEF and aEF paths lie significantly higher in the outer region. Consequently, the spontaneous fission in this nucleus is expected to proceed along the symmetric sCF path corresponding to the higher-TKE mode. This transition from an asymmetric fission path in $^{256}$Fm to a compact-symmetric path in $^{260}$Fm is due to shell effects in the emerging fission fragments approaching the doubly magic nucleus $^{132}$Sn (see, e.g., discussion in Refs. [18, 19, 20]).

## 4. Summary

We analyzed the static spontaneous-fission valleys in $^{256,258,260}$Fm within the self-consistent SHF+BCS framework. For all the nuclei investigated, we found the symmetric-compact, symmetric-elongated, and asymmetric-elongated static paths to fission, leading to symmetric-spherical, symmetric-elongated, and asymmetric-elongated fragment splits at the scission point, respectively.

In the $^{256}$Fm isotope, we found the asymmetric (aEF) path to be most favorable. In the case of $^{258}$Fm, the two symmetric (sCF and sEF) paths take over. This, together with a small asymmetric contribution from the aEF path, explains the bimodal fission observed in this isotope. For the heaviest $^{260}$Fm isotope, we found a purely symmetric fission leading to nearly spherical scission fragments. To what extent will dynamical effects [8] modify this picture? The answer to this question will be provided in forthcoming studies.

## Acknowledgements

This work was supported in part by the National Nuclear Security Administration under the Stewardship Science Academic Alliances program through the U.S. Department of Energy Research Grant DE-FG03-03NA00083; by the U.S. Department of Energy under Contract Nos. DE-FG02-96ER40963 (University of Tennessee), DE-AC05-00OR22725 with UT-Battelle, LLC (Oak Ridge National Laboratory), and DE-FG05-87ER40361 (Joint Institute for Heavy Ion Research); by the Polish Committee for Scientific Research (KBN) under Contract No. 1 P03B 059 27; by the Foundation for Polish Science (FNP); and by the Polish Ministry of Science and Higher Education under Contract No. N202 179 31/3920.